\documentclass[a4paper,11pt]{article}
\usepackage{jheppub}

\makeatletter
\gdef\@fpheader{}
\makeatother

\usepackage{amsmath,amssymb,amsfonts}
\usepackage{mathtools}
\usepackage{bm}
\usepackage{mathrsfs}
\usepackage{lmodern}
\usepackage{simpler-wick}
\usepackage{subcaption}
\usepackage[labelfont=bf, margin=20pt, font={small}]{caption}
\usepackage[utf8]{inputenc}
\usepackage[T1]{fontenc}
\usepackage[english]{babel}
\usepackage{graphicx}
\usepackage{braket}
\usepackage{textcomp}
\usepackage{slashed}
\usepackage[dvipsnames,svgnames,table]{xcolor}
\usepackage{tikz}
\usepackage{tikz-cd}
\usepackage[normalem]{ulem}
\usepackage{float}
\usepackage{multirow}
\usepackage[makeroom]{cancel}

\graphicspath{{figures/}}

\hypersetup{
    colorlinks=true,
    linktocpage,
    breaklinks,
    urlcolor=blue,
    linkcolor=blue,
    citecolor=blue
}

\newcommand\beq{\begin{equation}}
\newcommand\eeq{\end{equation}}

\numberwithin{equation}{section}

\title{\boldmath Solutions in $\sqrt{\textbf{Liouville\;theory}}$ on $\textbf{dS}_D$ and $\textbf{AdS}_D$ backgrounds}

\author[a,b]{Damir Sadekov}

\affiliation[a]{Moscow Institute of Physics and Technology, Laboratory of High Energy Physics, Institutskii per., 9, 141702, Dolgoprudny, Russia}
\affiliation[b]{P. N. Lebedev Physical Institute, Moscow 119991, Russia}

\emailAdd{sadekov.di@phystech.edu}

\abstract{We find exact solutions of a Liouville-type scalar field theory on $D$-dimensional de Sitter and anti-de Sitter backgrounds, treating the geometry as nondynamical. Using the embedding-space formalism with an auxiliary null vector, we derive first-order (B\"acklund-like) equations whose integrability yields the solution of the Liouville equation for the scalar field. This method produces two classes of analytic solutions, whose physical properties are different for de Sitter and anti-de Sitter spacetimes.}

\begin{document}
\vspace*{0.5cm}
\maketitle
\flushbottom

\section{Introduction}

Quantum field theory in curved spacetime remains one of the most conceptually rich and actively developing areas of modern theoretical physics. Despite significant progress, it still contains numerous open and unresolved problems. This situation motivates the study of simple and analytically tractable models, which allow one to explore various aspects of quantum fields in nontrivial geometries in a controlled way. 

Among the simplest yet nontrivial curved backgrounds are de Sitter (dS) and anti-de Sitter (AdS) spacetimes, which are maximally symmetric spaces of constant curvature. De Sitter space plays a central role in modern cosmology: it provides a good approximation to the inflationary stage of the early Universe and to the present-day accelerated expansion driven by dark energy. On the other hand, anti-de Sitter space naturally arises in the context of the AdS/CFT correspondence and is of fundamental importance for the study of quantum gravity and strongly coupled quantum field theories.

At the same time, quantum field theory in both dS and AdS spacetimes exhibits a number of subtle and still not fully understood features. In de Sitter space, the presence of a cosmological horizon and the absence of a globally defined timelike Killing vector lead to ambiguities in the definition of vacuum states and particles. Moreover, interacting quantum fields in dS space are known to exhibit infrared (IR) effects, such as secular growth, which can significantly modify the late-time behavior of correlation functions \cite{Polyakov:2012uc, Krotov:2010ma, Akhmedov:2013vka, Akhmedov:2017ooy, Akhmedov:2019cfd, Moreau:2020gib, Gautier:2013aoa, Guilleux:2016oqv, Serreau:2013psa}. These effects are closely related to the physics of inflation and may have observable consequences, as well as conceptual implications for the stability of dS space and the resummation of quantum corrections \cite{Glavan:2021adm, Miao:2024nsz, Woodard:2025smz, Akhmedov:2013vka, Akhmedov:2024npw}. In AdS space, although the presence of a timelike boundary allows for a more conventional definition of conserved quantities, the theory is highly sensitive to boundary conditions imposed at spatial infinity. These boundary conditions play a crucial role in defining the spectrum of states and the dynamics of interacting fields, and are tightly connected to holography and the AdS/CFT correspondence. At the same time, interacting quantum fields in AdS also exhibit nontrivial ultraviolet and infrared features, as well as dependence on the choice of coordinate patches and global structure of spacetime \cite{Akhmedov:2018lkp, Akhmedov:2020jsi, Moschella:2025lqy}. Given these challenges, the study of simple but nontrivial models admitting analytic or semi-analytic solutions becomes particularly important.

The analysis of soliton solutions and other nonperturbative configurations on curved backgrounds is of considerable interest in this context. In contrast to flat spacetime, where solitons are typically stable and freely propagating objects, the presence of curvature can drastically affect their dynamics, stability, and interactions. In cosmological settings with an expanding dS metric, long-lived localized configurations such as oscillons have been extensively studied \cite{Amin:2010jq, Lozanov:2014zfa, Levkov:2023ncb, vanDissel:2025xqn}. These objects may play a role in post-inflationary dynamics and have been discussed as potential sources of gravitational waves or as dark matter candidates. Soliton-like configurations in curved spacetimes also appear in other contexts, such as cosmic strings and nonlinear sigma-models \cite{dePaduaSantos:2015oxy}. Finally, the recently discovered soliton solutions in sine-Gordon type theory on curved backgrounds \cite{Akhmedov:2026msi} provide a strong motivation to further investigate integrable or quasi-integrable structures in spaces of constant curvature. In the latter case, gravity remains nondynamical, i.e., the AdS and dS backgrounds remain intact, as is the case for integrable field theories on a flat background.

In this work, we consider a scalar field theory in spacetime dimensions $D \geq 2$ with an action of the Liouville form:
\beq\label{eq:liouville_action}
    S = \int d^{D}x \sqrt{|g|} \left[ \pm \frac{1}{2} g^{\alpha\beta} \partial_{\alpha}\varphi \partial_{\beta}\varphi - \frac{\mu^2}{2} e^{2a\varphi} - Q \mathcal{R} \varphi \right],
\eeq
where $\mathcal{R}$ is the Ricci scalar and the parameter $Q$ will be specified in Section \ref{sec:general_solution} by fixing the explicit potential $V[\varphi]$ that we consider. For a stable theory (assuming the mostly plus signature with negative $g_{00}$), the kinetic term should enter with a negative sign. However, the solutions that we consider in $\text{dS}_D$ correspond to the opposite sign, which formally leads to an instability. Such a situation can nevertheless arise in effective field theories because, for instance, the conformal factor of the metric in gravity is known to have a kinetic term with the ``ghost'' sign. In the present work, we will not address this issue in detail and will instead focus on classical solutions.

Liouville-type potentials in higher dimensions appear in several contexts, including higher-dimensional conformal field theories \cite{Levy:2018bdc, Gaikwad:2023gef} and gravitational solutions in Einstein--Maxwell--Dilaton theories \cite{Charmousis:2009xr}. In two dimensions, Liouville theory plays a fundamental role: it provides a nontrivial example of an exactly solvable interacting quantum field theory \cite{Seiberg:1990eb, kupiainen2020integrability} and a distinguished conformal field theory \cite{Teschner:2001rv}, is central to two-dimensional quantum gravity \cite{Distler:1988jt, Knizhnik:1988ak, duplantier2014liouville}, has also been extensively studied on hyperbolic geometries \cite{Zamolodchikov:2001ah}, and continues to be studied, for instance in the computation of higher-point correlation functions \cite{Artemev:2022nvl, Artemev:2024rck}.

At the classical level, the integrability of the Liouville equation manifests itself through its reduction to a system of first-order differential equations via B\"acklund transformations \cite{Rogers_1982}:
\beq\label{eq:flat_Backlund}
\begin{cases}
    \partial_{+}(u+v) = \lambda \mu\, e^{a(u-v)}\;, \\
    \partial_{-}(u-v) = \frac{1}{\lambda}\mu\, e^{a(u+v)} 
\end{cases}
\;\; \Longrightarrow \;\;\;\;
\begin{cases}
    \partial_{+}\partial_{-}u = \mu^2 e^{2au}\;, \\
    \partial_{+}\partial_{-}v = 0\;, 
\end{cases}
\eeq
where $\lambda$ is a free parameter and $\partial_{\pm}$ denote derivatives with respect to light-cone coordinates. This formulation shows that the general solution of the Liouville equation can be constructed from solutions of the free wave equation. In this paper, we develop a similar construction for $\text{dS}_D$ and $\text{AdS}_D$ backgrounds. The resulting ``B\"acklund-like'' transformations involve additional vector structures and lead to some nontrivial solutions.

The paper is organized as follows. In Section \ref{sec:general_solution}, we describe the general construction of the B\"acklund-like transformations in spaces of constant curvature. In Sections \ref{sec:dS} and \ref{sec:AdS}, we present explicit realizations of these solutions in $\text{dS}_D$ and $\text{AdS}_D$, respectively. Finally, in Section \ref{sec:discussion}, we summarize our results and discuss possible directions for future research.

\section{General construction of solutions}\label{sec:general_solution}
In the embedding space formalism, the $\text{dS}_D$ and $\text{AdS}_D$ spaces with curvature radius $R$ can be realized as hyperboloids embedded in $(D+1)$-dimensional flat spaces:
\beq\label{eq:dS_definition}
    \text{dS}_D\; = \;
    \left\{\; X^A\in \mathbb{R}^{D+1} \;\Big|
        \;X\cdot X \equiv \eta^{\text{dS}}_{AB}X^A X^B \equiv -X_0^2 + \sum\limits_{I=1}^{D}X_{I}^2 = R^2\;
    \right\},
\eeq
\beq\label{eq:AdS_definition}
    \text{AdS}_D\; = \;
    \left\{\;
        X^A\in \mathbb{R}^{D+1} \;\Big|\;X\cdot X \equiv \eta^{\text{AdS}}_{AB}X^A X^B \equiv -X_0^2 + \sum\limits_{I=1}^{D-1}X_{I}^2 - X_{D}^2 = -R^2\;
    \right\}.
\eeq
The covariant derivative and the d'Alembert operator on these spaces can be written in the following general form:
\beq\label{eq:covariant_deriv_definition}
\begin{aligned}
    \nabla_A = \partial_A - \frac{1}{X\cdot X} X_A X^B \partial_B\;, 
    \\
    \Box = \nabla^A \nabla_A = \partial^A\partial_A - \frac{1}{X\cdot X}\; \left( X^A\partial_A \right)^2 - \frac{D-1}{X\cdot X}\; X^A \partial_A\;,
\end{aligned}
\eeq
with the restrictions $X\cdot X = \pm R^2$. For future convenience, let us introduce the operator
\beq\label{eq:M_operator}
    M_{AB} = \frac{1}{R}\left(X_A\partial_B - X_B\partial_A\right),
\eeq
which generates the Lie algebra of the isometry group of the corresponding space and the Lorentz symmetry of the ambient space. One can then easily verify that the corresponding Casimir operator is proportional to the d'Alembert operator, with opposite signs in $\text{dS}_D$ and $\text{AdS}_D$:
\beq\label{eq:M_AB_squared}
    M^{AB}M_{AB} = 2\,\Box_{\text{dS}}\;,
    \quad
    M^{AB}M_{AB} = -2\,\Box_{\text{AdS}}\;,
\eeq
where contractions are performed using the metrics $\eta_{AB}^{\text{dS}}$ and $\eta_{AB}^{\text{AdS}}$, respectively.

It turns out that there exists a particularly useful variable in hyperbolic spaces, which allows one to simplify the equations of motion and construct a class of particular (though not general) solutions. This variable is defined in terms of a null vector $\xi^A$ in the embedding space, and is commonly used to introduce a plane-wave solutions in hyperbolic spaces \cite{Moschella:2025lqy, Bros:1995js, Akhmedov:2026msi}:
\beq\label{eq:Z_introduction}
    Z \equiv \frac{\xi^A X_A}{R}\;,\quad \xi^A\xi_A = 0\;.
\eeq
For convenience, we also introduce the tensor
\beq
    \xi_{AB} \equiv \frac{X_A\xi_B - X_B \xi_A}{\xi\cdot X}\;,
\eeq
which satisfies the following relations in both spacetimes under consideration, as can be verified by straightforward computation:
\beq\label{eq:xi_AB_relations}
    \xi_{AB} \xi^{AB} = -2\;,\quad M^{AB}\xi_{AB} = -2\;\frac{D-1}{R}\;. 
\eeq
Let us now consider the following system of first-order differential equations for two scalar fields $\phi_{+}$ and $\phi_{-}$ defined on these hyperbolic spaces:
\beq\label{eq:system}
    \begin{cases}
        M_{AB}\;\phi_{+} = \Big[\lambda\mu\; e^{a\phi_{-}} - \frac{D-1}{aR}\Big]\xi_{AB}\;,
        \\
        M_{AB}\;\phi_{-} = \Big[\frac{1}{\lambda}\mu\; e^{a\phi_{+}} - \frac{D-1}{aR}\Big]\xi_{AB}\;
    \end{cases}
\eeq
with an arbitrary real parameter $\lambda$. Acting with the operator $M_{AB}$ on the system (\ref{eq:system}) and using the relations (\ref{eq:xi_AB_relations}), we obtain the following second-order equations of motion in $\text{dS}_D$ and $\text{AdS}_D$ for the fields $\varphi = \frac{\phi_+ + \phi_-}{2}$ and $v = \phi_+ - \phi_-$:
\beq\label{eq:the_second_order_system_ds_and_ads}
    \begin{cases}
        \Box_{\text{dS}}\;\varphi = -a\mu^2\;e^{2a\varphi} + \frac{\left(D-1\right)^2}{aR^2}\;,
        \\
        \Box_{\text{dS}}\;v = 0\;;
    \end{cases}
    \quad
    \begin{cases}
        -\Box_{\text{AdS}}\;\varphi = -a\mu^2\;e^{2a\varphi} + \frac{\left(D-1\right)^2}{aR^2}\;,
        \\
        \Box_{\text{AdS}}\;v = 0\;.
    \end{cases}
\eeq
This shows that, in close analogy with the B\"acklund transformations (\ref{eq:flat_Backlund}), solutions of the first-order system (\ref{eq:system}) automatically provide solutions to the Liouville-type and free-wave equations (\ref{eq:the_second_order_system_ds_and_ads}) in curved backgrounds. As mentioned in the Introduction, the equation in $\text{dS}_D$ corresponds to a wrong-sign (ghost-like) kinetic term and is therefore formally unstable, whereas in $\text{AdS}_D$ one obtains an equation with the appropriate sign and a stable potential $V[\varphi]$, which includes the Liouville exponent and the coupling to the curvature:
\beq\label{eq:potential}
    V[\varphi] = \frac{\mu^2}{2}e^{2a\varphi} - \frac{\left(D-1\right)^2}{aR^2}\varphi\;.
\eeq
This potential grows exponentially for large positive $\varphi$ and linearly for large negative $\varphi$, and possesses a minimum at
\beq\label{eq:vacuum_value}
    \varphi_{\text{vac}} = \frac{1}{a}\log\left[\frac{D-1}{a\mu R}\right]\;, 
    \quad
    \gamma \equiv e^{a\varphi_{\text{vac}}}\;.
\eeq
A crucial difference from the standard B\"acklund transformations (\ref{eq:flat_Backlund}) is that the relations (\ref{eq:system}) involve a tensor structure determined by the external vector $\xi^A$. It restricts the class of admissible solutions, because we must satisfy this structure. To do this, we assume that the fields $\phi_{\pm}$ depend only on the variable $Z$ introduced in (\ref{eq:Z_introduction}), i.e. $\phi_{\pm} = \phi_{\pm}(Z)$. This ansatz is compatible with the tensor structure and reduces the system (\ref{eq:system}) to
\beq\label{eq:system_with_Z}
    \begin{cases}
        Z\partial_Z\phi_{+} = \lambda\mu\; e^{a\phi_{-}} - \frac{D-1}{aR}\;,
        \\
        Z\partial_Z\phi_{-} = \frac{1}{\lambda}\mu\; e^{a\phi_{+}} - \frac{D-1}{aR}\;.
    \end{cases}
\eeq
In this case, the second-order equations (\ref{eq:the_second_order_system_ds_and_ads}) take the form
\beq\label{eq:2_order_system_with_Z}
    \begin{cases}
        -Z^2\partial^2_Z\varphi-DZ\partial_Z\varphi = -a\mu^2\;e^{2a\varphi} + \frac{\left(D-1\right)^2}{aR^2}\;,
        \\
        -Z^2\partial^2_Zv-DZ\partial_Zv  = 0\;;
    \end{cases}
    \quad
    v(Z) = c_1+\frac{c_2}{Z^{D-1}}\;,
\eeq
where we also write down the solution of the free-wave equation for $v(Z)$. Thus, the solutions of (\ref{eq:the_second_order_system_ds_and_ads}) have the same functional form in $\text{dS}_D$ and $\text{AdS}_D$ when expressed in terms of $Z$, although the geometric realization of $Z$ differs in the two cases. This will be discussed in the following sections.

The system (\ref{eq:system_with_Z}) possesses a first integral $C$, which can be written as
\beq\label{eq:first_integral}
    e^{a\phi_{+}}-\lambda^2 e^{a\phi_-} = \frac{C}{|Z|^{D-1}}\;.
\eeq
We emphasize that the equations (\ref{eq:system_with_Z})--(\ref{eq:2_order_system_with_Z}) are invariant under rescalings of $Z$, including $Z\rightarrow -Z$. Therefore, one of the integration constants can always be absorbed into $Z$, and it is convenient to work with $|Z|$. The matching conditions at $Z=0$ for the second-order equations (\ref{eq:2_order_system_with_Z}) are satisfied even if the first derivatives of the fields are discontinuous, due to the overall factor of $Z^2$ multiplying the second derivatives. 

Next, expressing $\phi_{-}$ in terms of $\phi_{+}$ using (\ref{eq:first_integral}) and substituting into (\ref{eq:system_with_Z}), one obtains a solvable differential equation for $\phi_{+}$. The resulting solutions are
\beq\label{eq:solutions_for_phi_pm}
    \begin{aligned}
        \phi_+(Z) = \frac{1}{a}\log\left[\frac{C}{|Z|^{D-1}}\frac{\exp\left(\frac{1}{\gamma}\frac{C}{\lambda}\frac{1}{|Z|^{D-1}}\right)}{\exp\left(\frac{1}{\gamma}\frac{C}{\lambda}\frac{1}{|Z|^{D-1}} \right)+K}\right]\;,
        \\
        \phi_-(Z) = \frac{1}{a}\log\left[\frac{C}{\lambda^2|Z|^{D-1}}\frac{-K}{\exp\left(\frac{1}{\gamma}\frac{C}{\lambda}\frac{1}{|Z|^{D-1}} \right)+K}\right]\;,
    \end{aligned}
\eeq
where $K$ is an integration constant. These expressions are well-defined for all nonzero $Z$ provided that $C>0,\;\lambda>0,\;K\in[-1,0)$. From these expressions, one immediately obtains the corresponding solutions of the Liouville-type and free-wave equations:
\beq\label{eq:liouville_&_wave_general_sollutions}
    \begin{aligned}
        \varphi(Z) = \frac{\phi_+ + \phi_-}{2} = 
        -\frac{1}{a}\log\Bigg[\frac{|Z|^{D-1}}{A}\;\sinh\left(\frac{1}{\gamma}\frac{A}{|Z|^{D-1}}-\frac{1}{2}\log\left(-K\right)\right)\Bigg]\;\;,\;\; A\equiv\frac{C}{2\lambda}\;,
        \\
        v(Z)=\phi_+ - \phi_- = \frac{1}{a}\log\left(\frac{\lambda^2}{-K}\right)\;+\;\frac{C}{a\gamma\lambda}\frac{1}{|Z|^{D-1}}\;.
    \end{aligned}
\eeq

In what follows, we will not further consider the free-wave solution $v(Z)$. For generic values of $K$, the solution $\varphi(Z)$ is ill-defined at $Z=0$ and diverges as $|Z|\rightarrow\infty$. However, there are two special choices that lead to physically meaningful solutions:
\beq\label{eq:meaningful_solution_1}
        \varphi_1(Z) =  -\frac{1}{a}\log\Bigg[\frac{|Z|^{D-1}}{A}\;\sinh\left(e^{-a\varphi_{\text{vac}}}\;\frac{A}{|Z|^{D-1}}\right)\Bigg]\;,
\eeq
\beq\label{eq:meaningful_solution_2}
\varphi_2(Z) = \varphi_{\text{vac}}\;-\; \frac{1}{a}\log\Big[1+b\cdot|Z|^{D-1}\Big]\;,\;b>0\;.
\eeq
The first solution corresponds to the choice $K=-1$, while the second can be obtained either directly from (\ref{eq:system_with_Z}) for $C=0$, or as the limit $C\rightarrow0$ of (\ref{eq:liouville_&_wave_general_sollutions}) with $K = -1+\frac{C}{\lambda\gamma}b$, where $b>0$ is a finite integration constant. 

The function $\varphi_1(Z)$ is undefined at $Z=0$ and approaches $\varphi_{\text{vac}}$ as $|Z|\rightarrow\infty$, whereas $\varphi_2(Z)$ approaches $\varphi_{\text{vac}}$ at $Z=0$ and diverges as $|Z|\rightarrow\infty$. It is also worth noting that in the formal flat-space limit $R\rightarrow\infty$, the potential $V[\varphi]$ tends to a single Liouville exponent, while $\varphi_{\text{vac}}\rightarrow -\infty$ and the solutions~(\ref{eq:meaningful_solution_1})--(\ref{eq:meaningful_solution_2}) tend to minus infinity, thus formally solving the ordinary Liouville equation of motion in this limit. Hence, the solutions~(\ref{eq:meaningful_solution_1})--(\ref{eq:meaningful_solution_2}) are characteristic of curved backgrounds. In the following sections, we construct explicit realizations of these solutions in different patches of $\text{dS}_D$ and $\text{AdS}_D$ spacetimes.

\section{Realizations in de Sitter space}\label{sec:dS}
The $\text{dS}_D$ spacetime (\ref{eq:dS_definition}) in global coordinates, together with a general null vector $\xi^A$ in the ambient space, can be parametrized as follows:
\beq\label{eq:X_&_xi_parametrization_global_dS}
    \begin{cases}
        X_0 = R\sinh\left(\frac{t}{R}\right)\;,
        \\
        X_{I} = R\cosh\left(\frac{t}{R}\right)n_{I}\;,\;I=1,\ldots,D\;,
    \end{cases}
    \quad
    \xi^A = 
    \begin{pmatrix}
        1
        \\
        \bm{q}
    \end{pmatrix},
\eeq
with $|\bm{n}|=|\bm{q}|=1$. The time coordinate runs over $t\in(-\infty,+\infty)$, while the spatial slices are $(D-1)$-dimensional spheres $\bm{n}\in\mathbb{S}_{D-1}$.

\begin{figure}[t]
    \centering
    \def\svgwidth{\textwidth}
    \includegraphics[width=\linewidth]{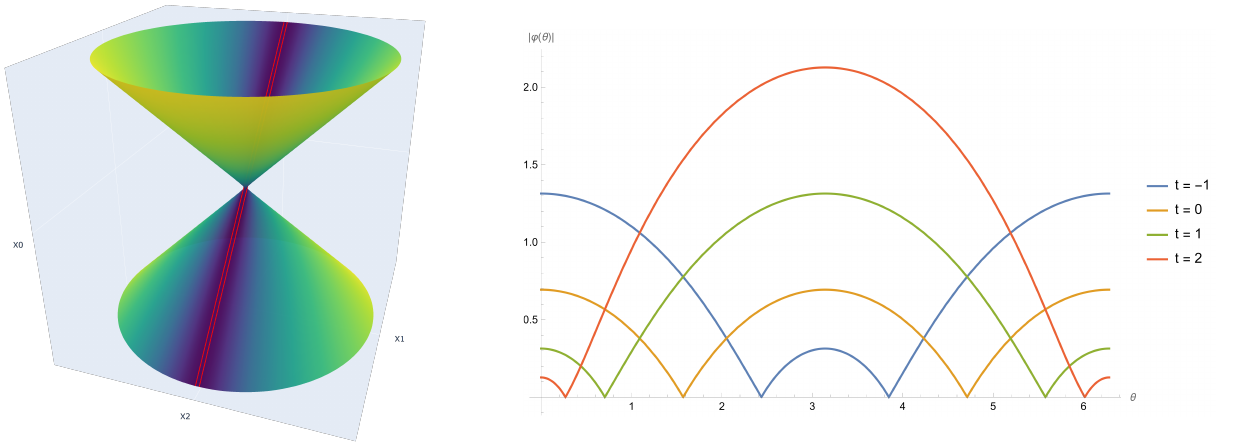}
    \caption{The left figure shows the absolute value of the solution $\varphi_2$ on the $\text{dS}_2$ background. Low values correspond to dark blue, while high values correspond to yellow. The red lines depict the points where the field equals $\varphi_{\text{vac}}$. The right figure shows the distribution of $|\varphi_2(\theta)|$ on the spatial circle at different moments of time.}
    \label{Fig:ds2}

    \vspace{1em}

    \centering
    \def\svgwidth{\textwidth}
    \includegraphics[width=\linewidth]{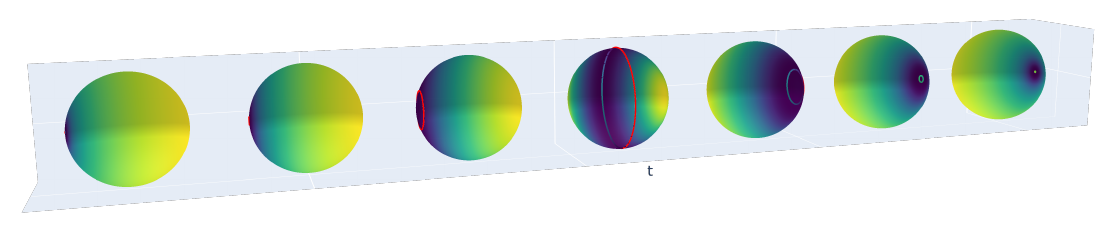}
    \caption{The distribution of $|\varphi_2|$ on the spatial spheres of $\text{dS}_3$ for $t=-3,-2,-1,0,1,2,3$ (in units of $R$). The vacuum value (red line) propagates from one pole of the sphere to the other.}
    \label{Fig:ds3}
    \hfill    
\end{figure}

In global coordinates, at any finite time $t$ there exists a set of points on $\mathbb{S}_{D-1}$ where $Z=0$:
\beq\label{eq:Z=0_condition_global_dS}
    Z=0 \quad
    \Longleftrightarrow
    \quad
    \bm{n}\cdot\bm{q} = \tanh\left(\frac{t}{R}\right).
\eeq
Therefore, the solution (\ref{eq:meaningful_solution_1}) is not well-defined globally in $\text{dS}_D$. The second solution (\ref{eq:meaningful_solution_2}) diverges at large times $t\rightarrow \pm \infty$, since in this limit $|Z|$ grows without bound:
\beq\label{eq:phi_1_divergence_global_dS}
    |Z|\sim 
    \begin{cases}
        \frac{1}{2}e^{t/R}\;\Big[1-\bm{n}\cdot\bm{q}\Big]\;,\;\text{as}\;t\rightarrow+\infty\;,
        \\
        \frac{1}{2}e^{-t/R}\;\Big[1+\bm{n}\cdot\bm{q}\Big]\;,\;\text{as}\;t\rightarrow-\infty\;,
    \end{cases}
\eeq
except at special points on the boundary, where $\bm{n}\cdot\bm{q}=\pm1$ at future and past infinity, respectively. At these points one has $Z\rightarrow 0$, and the solution approaches $\varphi_{\text{vac}}$. Thus, the solution can be interpreted as a ``flow'' of the vacuum value of the field from one pole of the spatial sphere $\mathbb{S}_{D-1}$ to the opposite pole. This behavior is illustrated in Fig.~\ref{Fig:ds2} and Fig.~\ref{Fig:ds3} for the cases of $\text{dS}_2$ and $\text{dS}_3$ \cite{dasadekov2026code}. There we plot the absolute value $|\varphi_2(Z)|$, which qualitatively reflects the energy density: for large $|Z|$, the field $\varphi_2$ becomes negative, and the stress-energy tensor is dominated by the potential term $V[\phi_2]\sim -\frac{(D-1)^2}{aR^2}\varphi_2$. The figures demonstrate how the region where the field is close to $\varphi_{\text{vac}}$ propagates across the sphere, while the magnitude of the field increases away from this region.

Although the solutions discussed above correspond to a formally unstable theory in de Sitter space, they still provide a useful qualitative picture of possible dynamical behavior. First, in a typical physical setup, one specifies initial data on a Cauchy surface at some finite time $t_0$, so that there is no issue with the growth of the fields at the asymptotic past, as they will start from some finite values at the Cauchy surface. Second, in a more realistic setting, the growth of the solution $\varphi_2(Z)$ with time would eventually be affected by gravitational backreaction or interactions with additional fields, which could spoil the unbounded increase of the field and modify the late-time behavior. Therefore, solutions of this type may serve as useful toy models for studying nontrivial time-dependent processes in dS backgrounds and motivate further investigation in more complete frameworks.

Let us now consider the Poincar\'e patch of $\text{dS}_D$ spacetime:
\beq\label{eq:poincare_coordinates}
    \begin{cases}
        X_0 = R\;\bigg[\sinh\left(\frac{\tau}{R}\right)+\frac{\bm{x}^2}{2R^2}e^{\frac{\tau}{R}}\bigg]\;,
        \\
        X_{I} = x_I\;,\;\;I=1,\ldots,D-1\;,
        \\
        X_D = R\;\bigg[-\cosh\left(\frac{\tau}{R}\right)+\frac{\bm{x}^2}{2R^2}e^{\frac{\tau}{R}}\bigg]\;,
    \end{cases}
    \quad
    \xi^A = 
    \begin{pmatrix}
        1
        \\
        \bm{q}_{\perp}
        \\
        q_D
    \end{pmatrix},
    \quad
    \bm{q}_{\perp}^{2} + q_{D}^{2}=1\;,
\eeq
where $\bm{x}\in\mathbb{R}_{D-1}$ and $\tau\in(-\infty,+\infty)$. In these coordinates, the variable $Z$ takes the form
\beq\label{eq:Z_in_poincare_dS}
    -Z = \sinh\left(\frac{\tau}{R}\right) - \frac{\bm{x}\cdot\bm{q}_{\perp}}{R}\;e^{\frac{\tau}{R}} + \frac{\bm{x}^2}{2R^2}\;e^{\frac{\tau}{R}}\left(1-q_D\right) + \cosh\left(\frac{\tau}{R}\right)q_{D}\;.
\eeq
In general, $|Z|$ grows at spatial infinity. However, for a special choice of the null vector, namely $\bm{q}_{\perp} = \bm{0}$ and $q_D = 1$, one finds that $|Z| = e^{\frac{\tau}{R}}$, so that the field depends only on time. In this homogeneous case, both solutions (\ref{eq:meaningful_solution_1}) and (\ref{eq:meaningful_solution_2}) can be realized:
\beq\label{eq:solutions_poincare_dS_homogeneous}
\begin{aligned}
    \varphi_1(\tau) = -\frac{1}{a}\log\Bigg[ e^{(D-1)\frac{\tau}{R}}\;\sinh\left(e^{-a\varphi_{\text{vac}}}\;e^{-(D-1)\frac{\tau}{R}}\right)\Bigg]\;,
    \\
    \varphi_2(\tau) = \varphi_{\text{vac}}\;-\; \frac{1}{a}\log\Big[1+e^{(D-1)\frac{\tau}{R}}\Big]\;.
\end{aligned}
\eeq
The first solution describes a homogeneous relaxation of the field toward the value $\varphi_{\text{vac}}$ from a large value at some initial moment $\tau_0$, while the second corresponds to a monotonic growth of the magnitude of the field starting from $\varphi_{\text{vac}}$ at past infinity. The analysis presented above demonstrates how the general construction is realized in dS space, and in the next section we turn to the corresponding realizations in $\text{AdS}_D$, where the structure of solutions exhibits important differences.

\section{Realizations in anti-de Sitter and hyperbolic spaces}\label{sec:AdS}
In two dimensions, $\text{AdS}$ and $\text{dS}$ spaces are very similar geometrically up to a rotation of the hyperboloid in the ambient space. In higher dimensions $D\geq 3$, however, $\text{AdS}_D$ possesses an additional nontrivial feature: there exists a vector space $\mathbb{V}_{\text{null}}$ of mutually orthogonal and non-coplanar null vectors \cite{Akhmedov:2026msi}:
\beq\label{eq:space_of_null_vectors}
    \forall \;\;i,j,\;\;\text{if}\quad \xi_{i}^{A},\;\xi_{j}^{A}\in \mathbb{V}_{\text{null}}\quad
    \Longrightarrow
    \quad
    \xi_i\cdot\xi_j = 0\;,\;\xi^2_i=\xi^2_j=0\;,
    \quad    
\eeq
This property provides additional freedom in constructing solutions. In particular, for two different vectors $\xi_i,\xi_j\in\mathbb{V}_{\text{null}}$, one finds relations analogous to (\ref{eq:xi_AB_relations}):
\beq
    (\xi_i)_{AB}(\xi_{j})^{AB} = \frac{X_A\xi_{iB}-X_B\xi_{iA}}{\xi_i\cdot X}\cdot
    \frac{X^A\xi_{j}^B-X^B\xi_{j}^A}{\xi_j\cdot X} = -2\;.
\eeq
This allows one to generalize the system (\ref{eq:system}), for example, to
\beq\label{eq:system_ads_multiple_xi}
    \begin{cases}
        M_{AB}\;\phi_{+} = \Big[\lambda\mu\; e^{a\phi_{-}} - \frac{D-1}{aR}\Big]\displaystyle\sum\limits_{i=1}^{N}k_i\cdot(\xi_i)_{AB}\;,
        \\
        M_{AB}\;\phi_{-} = \Big[\frac{1}{\lambda}\mu\; e^{a\phi_{+}} - \frac{D-1}{aR}\Big]\displaystyle\sum\limits_{i=1}^{N}k_i\cdot(\xi_i)_{AB}\;
    \end{cases}
\eeq
with $\xi_i\in\mathbb{V}_{\text{null}}$ and $k_i$ are arbitrary real constants satisfying $\sum k_i = 1$, so that this system still leads to the same equations (\ref{eq:the_second_order_system_ds_and_ads}) for $\text{AdS}_D$. From (\ref{eq:system_ads_multiple_xi}) it immediately follows that any solution (\ref{eq:liouville_&_wave_general_sollutions}) in $\text{AdS}_D$ can be generalized as
\beq\label{eq:ads_generalization_of_solution}
    \varphi(Z)\longrightarrow\varphi\left(\overline{Z}\right),\quad \overline{Z}=Z_1^{k_1}\cdot\ldots\cdot Z_N^{k_N},\;\;Z_i\equiv\xi_i\cdot X\;,\; R=1\;.
\eeq
For concreteness, let us consider the case of $\text{AdS}_3$. In global coordinates, the spacetime and the space $\mathbb{V}_{\text{null}}$ can be parametrized as
\beq\label{eq:ads3_parametrization}
    \begin{cases}
        X_0 = \cosh(\rho)\cos(t),
        \\
        X_1=\sinh(\rho)\sin(\Phi),
        \\
        X_2=\sinh(\rho)\cos(\Phi),
        \\
        X_3=\cosh(\rho)\sin(t);
    \end{cases}
    \quad
    \xi_i^A = a_i
    \begin{pmatrix}
        1\\
        1\\
        0\\
        0
    \end{pmatrix}
    + b_i
    \begin{pmatrix}
        0\\
        0\\
        1\\
        1
    \end{pmatrix}\;\equiv\; a_i\eta_a^A + b_i\eta_b^A\;\in\;\mathbb{V}_{\text{null}}\;,
\eeq
where $\rho\in[0,+\infty)$, $\Phi\in[0,2\pi)$, $t\in[0,2\pi)$, and $a_i,\;b_i$ are arbitrary real coefficients, while $\eta_a,\;\eta_b$ form a basis in $\mathbb{V}_{\text{null}}$. As in the $\text{dS}_D$ case, for any $\xi^A$ the variable $Z=\xi\cdot X$ vanishes at certain points in the bulk of global $\text{AdS}_D$. Therefore, the solution (\ref{eq:meaningful_solution_1}) with $Z\rightarrow \overline{Z}$ is not well-defined in this spacetime. 

\begin{figure}
    \centering
    \def\svgwidth{\textwidth}
    \includegraphics[width=\linewidth]{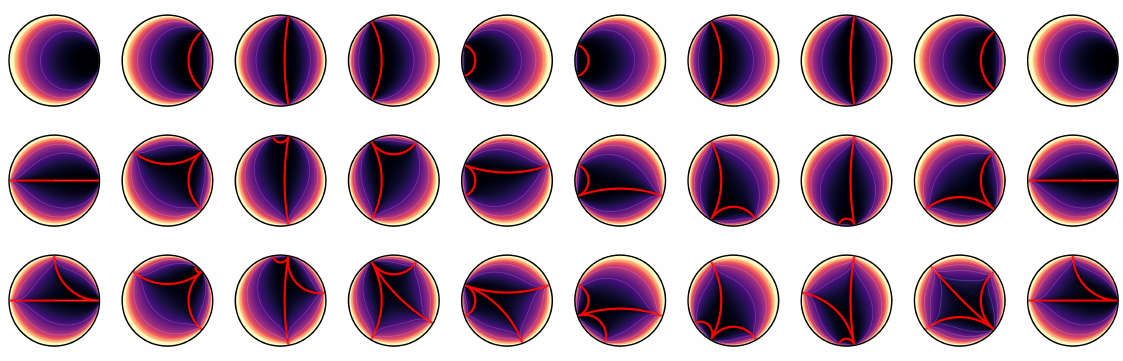}
    \caption{The plot of $|\varphi_2(\overline{Z})|$ in $\text{AdS}_3$ on the Poincar\'e disk for times from $t=0$ to $t=2\pi$. The panels correspond to $\overline{Z} = Z_1$, $\overline{Z}=(Z_1 Z_2)^{\frac{1}{2}}$, and $\overline{Z}=(Z_1 Z_2 Z_3)^{\frac{1}{3}}$, where $Z_1=\eta_a\cdot X,\; Z_2=\eta_b\cdot X,\;Z_3=(\eta_a+\eta_b)\cdot X$. The boundary is cut at $r=\tanh\left(\rho/2\right)=0.98$.}
    \label{Fig:ads3}  
\end{figure}

The second solution (\ref{eq:meaningful_solution_2}) diverges at spatial infinity $\rho\rightarrow +\infty$, except at special boundary points determined by the directions of $\xi_i\in\mathbb{V}_{\text{null}}$ entering (\ref{eq:ads_generalization_of_solution}). This behavior is illustrated in Fig.~\ref{Fig:ads3}, where we display the two-dimensional bulk of $\text{AdS}_3$ on the Poincar\'e disc, i.e. perform the coordinate transformation as follows:
\beq\label{eq:poincare_disc}
    \rho,\;\Phi\quad\longrightarrow\quad x=\tanh\left(\frac{\rho}{2}\right)\cos(\Phi),\;y=\tanh\left(\frac{\rho}{2}\right)\sin(\Phi)\;.
\eeq
The magnitude of the solutions interpolates between curves defined by $\overline{Z} = 0$ (shown in red), where the field approaches $\varphi_{\text{vac}}$. These curves intersect only at the spatial boundary and therefore cannot be interpreted as interacting solitons --- rather, they describe wave-like configurations propagating through the bulk without scattering.

Actually, even the consideration (\ref{eq:system_ads_multiple_xi})--(\ref{eq:ads_generalization_of_solution}) is redundant, and we have done it in order to show how to take the ``square root'' of the Liouville equation in another way in $\text{AdS}_D$. Indeed, if $\varphi=\varphi(Z_1,A)$ is a solution of some equation of motion of the form $\Box_{\text{AdS}}\;\varphi = F[\varphi]$ with a free parameter $A$, like integration constant, and $Z_1=\xi_1\cdot X$, then automatically
\beq\label{eq:general_phi_in_ads}
    \Box_{\text{AdS}}\;\varphi\left(Z_1, A\left(\frac{Z_1}{Z_2},\frac{Z_2}{Z_3},\ldots,\frac{Z_{N-1}}{Z_N}\right)\right)\;=\;F[\varphi] 
\eeq
for any number $N$ of variables $Z_i=\xi_i\cdot X$, provided $\xi_i\in\mathbb{V}_{\text{null}}$, and the free parameter $A$ becomes an arbitrary differentiable function of $N-1$ variables. This follows from the relations \cite{Akhmedov:2026msi}:
\beq\label{eq:relations_arbitrary_functions_of_ratios}
\begin{aligned}
    \Box_{\text{AdS}}\;A\left(\frac{Z_1}{Z_2},\frac{Z_2}{Z_3},\ldots,\frac{Z_{N-1}}{Z_N}\right) \;= 0\;,
    \\
    \nabla^AG(Z_1,\ldots,Z_N)\nabla_A A\left(\frac{Z_1}{Z_2},\frac{Z_2}{Z_3},\ldots,\frac{Z_{N-1}}{Z_N}\right)\;=\;0,
\end{aligned}
\eeq
for any function $G$. Thus, in field theories on $\text{AdS}_D$, one effectively acquires an additional ``free-wave'' degree of freedom, and the solution (\ref{eq:ads_generalization_of_solution}) represents just one realization of this freedom. In the flat-space limit $R\rightarrow \infty$ in higher than two dimensions, such functions of ratios reduce to standard traveling wave packets of the form $f(t-x)$ \cite{Akhmedov:2026msi}.

Let us now consider the Poincar\'e patch of $\text{AdS}_3$:
\beq\label{eq:poincare_coordinates_ads3}
    \begin{cases}
        X_0 = \frac{z}{2}\left[1+\frac{R^2+(x^2-t^2)}{z^2}\right]\;,
        \\
        X_1=\frac{z}{2}\left[1-\frac{R^2-(x^2-t^2)}{z^2}\right]\;,
        \\
        X_2=R\frac{x}{z}\;,
        \\
        X_3=R\frac{t}{z}\;.
    \end{cases}
\eeq
The situation is qualitatively similar to the Poincar\'e patch of $\text{dS}_D$: the solution (\ref{eq:meaningful_solution_2}) grows at large spatial coordinate $x$ (more precisely, at large $x^2-t^2$) for a general choice of null vectors. However, the special choice $\xi^A=\eta_a^A$ yields $Z=-\frac{R}{z}$, which depends only on the radial coordinate. In this case, the solution (\ref{eq:meaningful_solution_1}) describes a configuration approaching $\varphi_{\text{vac}}$ at the boundary $z=0$ and growing homogeneously into the bulk, while (\ref{eq:meaningful_solution_2}) requires introducing a cutoff $z_0>0$ and describes a decreasing profile starting from large values at $z=z_0$. These homogeneous radial solutions can be written in arbitrary spacetime dimension $D$:
\beq\label{eq:solutions_poincare_ads_homogeneous}
\begin{aligned}
    \varphi_1(z) = -\frac{1}{a}\log\Bigg[ \frac{R^{D-1}}{z^{D-1}}\;\sinh\left(e^{-a\varphi_{\text{vac}}}\;\frac{z^{D-1}}{R^{D-1}}\right)\Bigg]\;,
    \\
    \varphi_2(z) = \varphi_{\text{vac}}\;-\; \frac{1}{a}\log\left[1+\frac{R^{D-1}}{z^{D-1}}\right]\;,\;z>z_0\;.
\end{aligned}
\eeq
Let us also note that these solutions in $\text{AdS}_D$ are stable under small perturbations. Linearizing (\ref{eq:the_second_order_system_ds_and_ads}), one finds
\beq\label{eq:small_perturbation_ads3}
\begin{aligned}
    \varphi=\varphi_{1,2}(z)\;+\;e^{-i\omega t}z^{\frac{D-2}{2}}\;\chi(z)\quad\Longrightarrow
    \\
    \Longrightarrow\quad-\partial^2_z\chi(z) + \frac{1}{z^2}\Bigg[\frac{D(D-2)}{2}+2a^2\mu^2R^2\;e^{2a\varphi_{1,2}(z)}\Bigg]\chi(z)=\omega^2\chi(z)\;,
\end{aligned}
\eeq
which is a Schr\"odinger-type equation with a positive potential, implying a positive spectrum of $\omega$. Hence, such solutions can be viewed as non-vacuum stable background field configurations in some physical settings.

\begin{figure}
    \centering
    \def\svgwidth{\textwidth}
    \includegraphics[width=\linewidth]{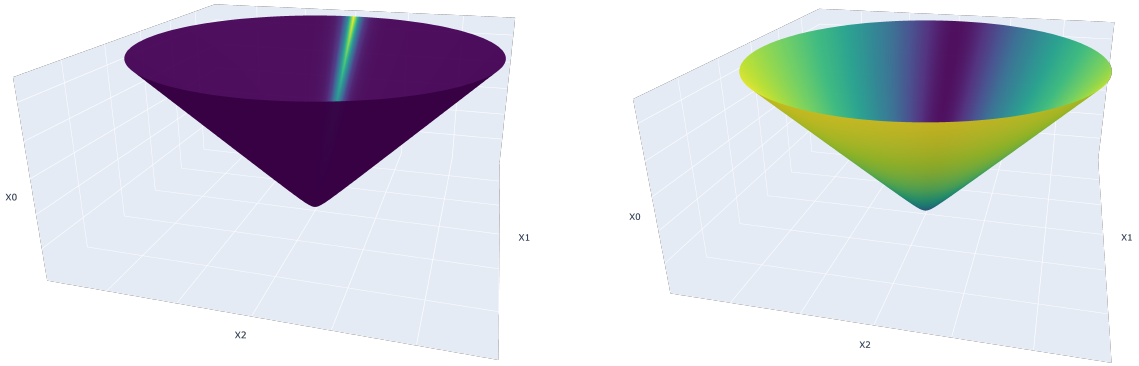}
    \caption{The visualization of the absolute value of the solutions $\varphi_{1,2}$ on the backgound of the hyperbolic space $\mathbb{H}_2$.}
    \label{Fig:h2}
\end{figure}

Finally, consider the hyperbolic Lobachevsky space $\mathbb{H}_2$, which can be viewed, for instance, as a spatial slice of global $\text{AdS}$. For illustration, we take the two-dimensional case:
\beq\label{eq:H2_definition}
    \begin{cases}
        X_0 = R\cosh(\psi)\;,
        \\
        X_1 = R\sinh(\psi)\cos(\theta)\;,
        \\
        X_2 =  R\sinh(\psi)\sin(\theta)\;,
    \end{cases}
    \quad
    \psi\in [0,\infty),\;\theta\in[0,2\pi),
    \quad
    \xi^A = 
    \begin{pmatrix}
    q
    \\
    q\cos(\alpha)
    \\
    q\sin(\alpha)
    \end{pmatrix}\;,
\eeq
which parametrizes the hyperboloid $-X_0^2+X_1^2+X_2^2 = -R^2$ together with a null vector. In this case, the variable $|Z|$ is always positive \cite{Akhmedov:2026msi}:
\beq\label{eq:H2_Z_variable}
    |Z| = \cosh(\psi) - \sinh(\psi)\cos(\theta-\alpha).
\eeq
At large $\psi$, it diverges except at the point $\theta=\alpha$, where it tends to zero. The solutions (\ref{eq:meaningful_solution_1})--(\ref{eq:meaningful_solution_2}) are shown in Fig.~\ref{Fig:h2}. The magnitude of $\varphi_1$ forms a beam-like profile aligned with the direction of $\xi^A$, rapidly decaying away from it, while $\varphi_2$ approaches $\varphi_{\text{vac}}$ along this direction and grows away from it as $\psi$ increases.
\section{Discussion}\label{sec:discussion}
In this work we have constructed a class of exact solutions of a Liouville-type scalar field theory on constant-curvature backgrounds, based on a first-order Bäcklund-like system formulated in the embedding space. Our primary motivation was to explore the possibility of extending structures reminiscent of integrability to quantum field theories in curved spacetime. In this context, the Bäcklund-like formulation can be viewed as a candidate mechanism for reducing nonlinear equations of motion to a system with auxiliary degrees of freedom, like some external null vectors $\xi^A$. However, in the present form, this construction generates only a restricted class of solutions rather than the general solution space, and therefore does not yet provide a full analogue of integrability or factorization as in two-dimensional Liouville theory. An interesting feature arises in AdS backgrounds, where the existence of a space of mutually orthogonal null vectors leads to an additional functional freedom in the solutions. This property is qualitatively similar to structures recently observed in modified sine-Gordon–type models in curved space \cite{Akhmedov:2026msi}. Nevertheless, this enhancement does not straightforwardly translate into integrability: in particular, it does not lead to a complete factorization of the dynamics or to a closed solution-generating algebra. Therefore, one of the main open problems is to identify a model in curved spacetime that admits a genuinely factorized description of the equations of motion, analogous to Bäcklund transformations, but involving more general configurations, potentially associated with non-orthogonal null vectors in the embedding space. Such a generalization could provide a more complete realization of integrable structures.

We believe that the construction presented here represents a step in this direction. Achieving further progress may require extending the class of admissible tensor structures, relaxing the constraints on the auxiliary vectors, or incorporating additional fields and interactions.

\section*{Acknowledgments}
The author is grateful to Dmitrii Diakonov for fruitful discussions on the topic. He would also like to thank E.T. Akhmedov and K. Gubarev for their discussions and careful reading of the paper. The work was supported by grant No.~26-12-00330  from the Russian Science
Foundation (RSF). 

\bibliography{literature}

\bibliographystyle{unsrt}
\end{document}